\documentclass[a4paper,twocolumn,aps,prd,superscriptaddress,showpacs,showkeys,amsmath,amssymb,floatfix,nofootinbib]{revtex4-1}
\usepackage{lmodern}

\usepackage[T1]{fontenc}
\usepackage[utf8]{inputenc}
\usepackage{color}
\usepackage{amsmath}
\usepackage[unicode=true,pdfusetitle,
 bookmarks=true,bookmarksnumbered=true,bookmarksopen=true,bookmarksopenlevel=1,
 breaklinks=false,pdfborder={0 0 0},pdfborderstyle={},backref=false,colorlinks=true]
 {hyperref}
\hypersetup{
 pdfborderstyle={},citecolor=blue,filecolor=blue,linkcolor=blue,urlcolor=blue}

\makeatletter

\pdfpageheight\paperheight
\pdfpagewidth\paperwidth

\usepackage{color}

\makeatother

\begin{document}

\title{Raychaudhuri equation in space-times with torsion}

\author{Paulo Luz}
\email{paulo.luz@ist.utl.pt}

\affiliation{Centro de Matemática, Universidade do Minho, Campus de Gualtar, 4710-057
Braga, Portugal}

\affiliation{Centro Multidisciplinar de Astrofísica - CENTRA, Departamento de
Física, Instituto Superior Técnico - IST, Universidade de Lisboa -
UL, Av. Rovisco Pais 1, 1049-001 Lisboa, Portugal}

\author{Vincenzo Vitagliano}
\email{vincenzo.vitagliano@ist.utl.pt}

\affiliation{Centro Multidisciplinar de Astrofísica - CENTRA, Departamento de
Física, Instituto Superior Técnico - IST, Universidade de Lisboa -
UL, Av. Rovisco Pais 1, 1049-001 Lisboa, Portugal}
\begin{abstract}
Given a space-time with non-vanishing torsion, we discuss the equation
for the evolution of the separation vector between infinitesimally
close curves in a congruence. We show that the presence of a torsion
field leads in general to tangent and orthogonal effects to the congruence;
in particular, the presence of a completely generic torsion field
contributes to a relative acceleration between test particles. We
derive, for the first time in the literature, the Raychaudhuri equation
for a congruence of time-like and null curves in a space-time with
the most generic torsion field. 
\end{abstract}
\maketitle

\section{Introduction\label{sec:Introduction}}

The appearance of singularities in a physical theory ineluctably marks
the pillars of Hercules of that model. General Relativity, from this
point of view, is not an exception: it did not take a long time for
the community of relativists to realize that even the two simplest
space-time solutions of Einstein field equations, the Schwarzschild
metric and the Friedmann–Lemaître–Robertson–Walker (FLRW) metric,
harbour a gravitational singularity, that is, a breakdown of the space-time
structure itself! While in the early 50's Amal Kumar Raychaudhuri
was forcedly working on the properties of electronic energy bands
in metals, he got interested in the debate around the nature of gravitational
singularities and the generic features of Einstein's theory of General
Relativity (GR). Motivated by cosmology, Raychaudhuri started from
the idea that a singularity is nothing more than an artifact of the
symmetries of the matter distribution; in his seminal paper \cite{Raychaudhuri}
he then proposed a model of time-dependent universe without assuming
the cosmological principle and its implications on homogeneity and
isotropy. The analysis of the flows kinematic that he carried out
resulted in the renown Raychaudhuri equation\footnote{Sometimes referred to as the Landau–Raychaudhuri equation.}
for the evolution of the cosmological expansion in a given background.
It was Raychaudhuri himself that pointed out the relation of his work
with the existence of singularities, and it should not come as a surprise
that Penrose and Hawking were late inspired by Raychaudhuri's paper
to define the conditions under which their singularity theorem holds
\cite{Penrose,Hawking}.

Since its inception, the Raychaudhuri equation and its subsequent
generalizations (see e.g. the nice review \cite{Sengupta}) have found
application not only, as already mentioned, in terms of singularity
theorems but also in a vast range of different physical contexts,
from the study of gravitational lensing \cite{Tomita}, to crack formation
in spherical astrophysical objects \cite{herrera}, and finally to
some more fundamental issues as in the case of the derivation of the
(modified) Einstein equations as equations of state for a (non) equilibrium
spacetime thermodynamics \cite{Jacobson,Eling_Guedens_Jacobson,chirco}.

In this paper we generalize, for the first time in the literature,
the Raychaudhuri construction to a congruence of curves embedded in
a space-time with a non-trivial, completely generic torsion tensor
field. The idea of generalizing Einstein General Relativity to non-Riemannian
geometries has been now around for a while. One of the prototypical
example of this kind of proposals is the Einstein–Cartan–Sciama–Kibble
(ECSK) theory. The ECSK theory is characterized by assuming an independent
connection (using the so-called Palatini approach to find two sets
of independent field equations) and further requiring the anti-symmetric
part of the connection to be in general non-vanishing, defining a
tensor field dubbed torsion tensor field; note, however, that the
compatibility of the connection with the metric is still imposed,
that is, the covariant derivative (defined with the independent connection)
of the metric tensor field along any space-time curve is null.

Theories of gravity with non-vanishing torsion have been extensively
considered in the literature; the attention toward these models has
been initially catalyzed by the relatively simple scheme they provide
to account for nontrivial quantum effects in a gravitational environment,
achieved through a direct coupling between the intrinsic spin of matter
and the torsion field \cite{Hehl,Trautman,Stewart,Kop,Tafel,sotiriou,enzo2,enzo3,Santana};
however this interest has been extended to the fact that torsion appears
naturally and inevitably in the low energy limit of super string theories,
any theory of gravity that considers twistors and in higher dimensional
Kaluza-Klein theories \cite{Capozziello}. On top of that, gravitational
theories with torsion exhibit a unique nature in comparison with other
modified theories of gravity: in general, unless some particular further
assumptions are imposed, the effects of a non-null torsion field cannot
be recast in an effective energy-momentum tensor, in other words,
the torsion field cannot be seen, in general, as an extra matter field
in a Lorentzian torsion free manifold. ECSK theory is a particular
exception to this property, where torsion can in fact be seen as the
effect of a matter field in a Lorentzian manifold, usually called
spin-torsion. Yet ECSK theory by no means represents the most generic
case of a theory equipped with a torsion field (see for example \cite{Mao,Capozziello})
and, although this circumstance is widely recognized, it is
intriguing to note that most of the work done on gravitational theories with non-vanishing torsion only considers the ECSK setup: for
instance, the very important question of whether the presence of torsion
can avoid the formation of singularities resulting from gravitational
collapse has only been answered for the case of a spin-torsion field
\cite{Trautman,Stewart,Kop}. In this paper we then start filling
such a crucial gap in the literature: we will focus on the study of
the effects of the most generic torsion field on the kinematics of
test particles and derive the Raychaudhuri equation for a congruence
of null and time-like curves in the spacetime.

The paper is organized as follows: in Sec.~\ref{sec:Conventions-and-Notations}
we introduce the basic definitions and set the conventions that will
be used throughout the article; in Sec.~\ref{sec:Raychaudhuri-Equation}
we derive the evolution equation of the separation vector between
test particles in space-times with torsion, define the kinematical
quantities of a congruence of curves and derive the Raychaudhuri equation,
both for a time-like and null congruence of curves; finally, we summarize
the main results and sort out our conclusions in Sec.~\ref{sec:Conclusion}.

\section{Conventions and Notations}

\label{sec:Conventions-and-Notations}

Due to the wide variety of different conventions used in literature,
let us start by gently introducing the basic definitions and setting
the conventions that will be used throughout the article. Introduce
a covariant derivative, 
\begin{equation}
\nabla_{\alpha}U^{\beta}=\partial_{\alpha}U^{\beta}+C_{\alpha\sigma}^{\beta}U^{\sigma}\,,\label{eq:Covariant_definition}
\end{equation}
constrained to be metric compatible, $\nabla_{\alpha}g_{\beta\gamma}=0$,
but otherwise with a completely generic connection $C_{\alpha\beta}^{\gamma}$.
The anti-symmetric part of the connection define a tensor which is
called the torsion tensor 
\begin{equation}
S_{\alpha\beta}\,^{\gamma}\equiv C_{\left[\alpha\beta\right]}^{\gamma}=\frac{1}{2}\left(C_{\alpha\beta}^{\gamma}-C_{\beta\alpha}^{\gamma}\right)\,.\label{eq:Torsion_tensor}
\end{equation}
Using such definition, it is possible to split the connection into
an appropriate combination of the torsion tensor plus the usual metric
Christoffel symbols $\Gamma_{\alpha\beta}^{\gamma}$, 
\begin{equation}
C_{\alpha\beta}^{\gamma}=\Gamma_{\alpha\beta}^{\gamma}+S_{\alpha\beta}\,^{\gamma}+S^{\gamma}\,_{\alpha\beta}-S_{\beta}\,^{\gamma}\,_{\alpha}\,.
\end{equation}
The sum of the three torsion pieces on the right hand side of last
equation is frequently dubbed in literature as the contorsion tensor,
$K_{\alpha\beta}\,^{\gamma}\equiv S_{\alpha\beta}\,^{\gamma}+S^{\gamma}\,_{\alpha\beta}-S_{\beta}\,^{\gamma}\,_{\alpha}$;
using the anti-symmetry of the torsion tensor, it is an easy task
to verify straightforwardly the two symmetries of contorsion, 
\begin{equation}
K_{\alpha\beta\gamma}=-K_{\alpha\gamma\beta}\,,\label{eq:Contorsion_last2indexes}
\end{equation}
\begin{equation}
K_{\left[\alpha\beta\right]}\,^{\gamma}=S_{\alpha\beta}\,^{\gamma}\,.\label{eq:Contorsion_first2indexes}
\end{equation}

Having in mind the general affine connection, the commutator between
two vectors can be expressed in terms of the torsion-full covariant
derivative as 
\begin{equation}
\left[u,v\right]^{\gamma}=u^{\alpha}\nabla_{\alpha}v^{\gamma}-v^{\alpha}\nabla_{\alpha}u^{\gamma}-2S_{\alpha\beta}\,^{\gamma}u^{\alpha}v^{\beta}\,.\label{eq:commutator_explicit}
\end{equation}
This last equation and the definition of the Riemann tensor associated
with $C_{\alpha\beta}^{\gamma}$, 
\begin{equation}
R_{\alpha\beta\gamma}\,^{\rho}=\partial_{\beta}C_{\alpha\gamma}^{\rho}-\partial_{\alpha}C_{\beta\gamma}^{\rho}+C_{\beta\sigma}^{\rho}C_{\alpha\gamma}^{\sigma}-C_{\alpha\sigma}^{\rho}C_{\beta\gamma}^{\sigma}\,.\label{eq:Riemann_tensor}
\end{equation}
lead to a modified version of the relation between the curvature and
the commutator of two covariant derivative in the case of non vanishing
torsion, 
\begin{equation}
R_{\alpha\beta\gamma}\,^{\rho}w_{\rho}=\left[\nabla_{\alpha},\nabla_{\beta}\right]w_{\gamma}+2S_{\alpha\beta}\,^{\rho}\nabla_{\rho}w_{\gamma}\,.\label{eq:Riemann_tensor_definition}
\end{equation}

The Riemann tensor for a non-symmetric connection does not have all
the usual symmetries. However, from \eqref{eq:Riemann_tensor} we
see that 
\begin{equation}
R_{\alpha\beta\gamma}\,^{\rho}=-R_{\beta\alpha\gamma}\,^{\rho}\,,
\end{equation}
and, using the symmetries of the contorsion tensor, \eqref{eq:Contorsion_last2indexes}
and \eqref{eq:Contorsion_first2indexes}, 
\begin{equation}
R_{\alpha\beta\gamma\rho}=-R_{\alpha\beta\rho\gamma}\,.\label{eq:Riemann_antisymmetry_last2indexes}
\end{equation}
So, the skew symmetry of the Riemann tensor is still verified. To
define the Ricci tensor in terms of the Riemann tensor we will adopt
the convention \cite{Enzo,enzobis} 
\begin{equation}
R_{\alpha\beta}=R_{\alpha\gamma\beta}\,^{\gamma}\,,
\end{equation}
which, using \eqref{eq:Riemann_tensor}, can be expressed in terms
of the connection coefficients as 
\begin{equation}
R_{\alpha\beta}=\partial_{\gamma}C_{\alpha\beta}^{\gamma}-\partial_{\alpha}C_{\gamma\beta}^{\gamma}+C_{\alpha\beta}^{\rho}C_{\gamma\rho}^{\gamma}-C_{\gamma\beta}^{\rho}C_{\alpha\rho}^{\gamma}\,.
\end{equation}
The Ricci scalar is defined as 
\begin{equation}
R=g^{\alpha\beta}R_{\alpha\beta}\,.
\end{equation}
Note that in the case of a general affine metric-compatible connection
it is still possible to consider an independent contraction of the Riemann
tensor defining a 2-rank tensor, $\widetilde{R}_{\alpha\beta}=g^{\gamma\delta}g_{\epsilon\beta}R_{\alpha\gamma\delta}\,^{\epsilon}$;
however a further contraction with the metric results in $\widetilde{R}=-R$,
that is the Ricci scalar is unequivocally defined.

A further important remark is about the notation we will be using
to describe the matter sector. For a general Riemann–Cartan theory
of gravity, the matter Lagrangian can couple (non-)minimally to the
torsion tensor which introduce new degrees of freedom in the problem.
While the stress-energy tensor is still defined as usual as the variation
of the matter action with respect to the metric, we need to introduce
a new object, the intrinsic hypermomentum, defined as the variation
of the action with respect to the independent connection, 
\begin{equation}
\Delta_{\mu\nu}{}^{\rho}\equiv\frac{1}{\sqrt{-g}}\frac{\delta S_{\textrm{Matter}}}{\delta\Gamma_{\mu\nu}{}^{\rho}}\,.
\end{equation}
Such quantity encapsulates all the information of the microscopic
structure of the particle, i.e. intrinsic spin, dilaton charge and
intrinsic shear.

\section{Raychaudhuri Equation\label{sec:Raychaudhuri-Equation}}

\subsection{The separation vector and its evolution}

Introduced the basic definitions and identities we are now in the
position to generalize the Raychaudhuri equation for the case of an
$N$-dimensional space-time with non-null torsion.

The notion of separation (sometimes deviation) vector between two
infinitesimally close curves is quite intuitive: define a congruence
of curves, not necessarily geodesics, such that each curve of the
congruence is parameterized by an affine parameter $\lambda$. Consider
a second congruence, this time of geodesics, parametrised by an affine
parameter $t$, such that each geodesic intersects a curve of the
first congruence at one and only one point of the space-time. Given
two curves in the first congruence, $c_{1}$ and $c_{2}$, and a geodesic
of the second congruence, $\gamma$, let the two points $p$ and $q$
be the intersection points of $\gamma$ with, respectively, $c_{1}$
and $c_{2}$, with $c_{1}\left(\lambda_{0}\right)=\gamma\left(t_{0}\right)=p$.
Let us now assume that the point $q$ is in a small enough neighbourhood
of the point $p$ such that $q=\gamma\left(t_{0}+\delta t\right)\approx p+\left.\frac{\partial\gamma}{\partial t}\right|_{t_{0}}\delta t$.
If $n$ is the tangent vector to the geodesic $\gamma$ in $p$, then
\begin{equation}
n\equiv\delta t\,\left.\frac{\partial\gamma}{\partial t}\right|_{t_{0}}=q-p
\end{equation}
gives also a meaningful notion of the separation between the curves
$c_{1}$ and $c_{2}$.

Let us consider a coordinate neighbourhood that contains the points
$p$ and $q$, such that $p=\left\{ x^{\alpha}\right\} $, $q=\left\{ x'^{\alpha}\right\} =\left\{ x^{\alpha}+n^{\alpha}\right\} $,
with 
\begin{equation}
n^{\alpha}=\delta t\frac{\partial x^{\alpha}}{\partial t}\,,\label{eq:separation_vector}
\end{equation}
and let $U^{\alpha}$ be the tangent vector to the curve $c_{1}$
(from here on we will drop the index $1$) at $p$, 
\begin{equation}
U^{\alpha}=\frac{\partial x^{\alpha}}{\partial\lambda}\,.\label{eq:tangent_vector}
\end{equation}
In order to find the general expression for the evolution of the separation
vector, $n^{\alpha}$, we will start by computing the Lie derivative
of $n$ over the tangent vector $U$ and vice versa. From the definition
of the Lie derivative and using \eqref{eq:separation_vector} and
\eqref{eq:tangent_vector} we find that 
\begin{equation}
\mathcal{L}_{n}\,U=\mathcal{L}_{U}\,n=0\,.\label{eq:n_v_lie_derivative}
\end{equation}
Using \eqref{eq:commutator_explicit} and \eqref{eq:n_v_lie_derivative}
it is possible to derive an equation for the change of the separation
vector along the fiducial curve $c$, 
\begin{equation}
U^{\beta}\nabla_{\beta}n^{\alpha}=B_{\beta}\,^{\alpha}n^{\beta}\,,\label{eq:derivative_n_relation}
\end{equation}
where 
\begin{equation}
B_{\beta}\,^{\alpha}=\nabla_{\beta}U^{\alpha}+2S_{\gamma\beta}\,^{\alpha}U^{\gamma}\,.\label{eq:B_tensor_general}
\end{equation}
For infinitesimally close curves, the evolution of the separation
vector along the fiducial curve is entirely described by the tensor
field $B_{\alpha\beta}$. Let us emphasize that in the derivation
of \eqref{eq:derivative_n_relation} and \eqref{eq:B_tensor_general}
we have not specified the type of the tangent vector to the fiducial
curve, $U^{\alpha}$, hence, this equations are equally valid for
the case of $U^{\alpha}$ being time-like, space-like or light-like,
with the fiducial curve being either a geodesic or not.

Let us now verify some physical implications of \eqref{eq:derivative_n_relation}
and \eqref{eq:B_tensor_general} in the case of the presence a non-vanishing
torsion tensor. The derivative along $c$ of the quantity $n_{\alpha}U^{\alpha}$
reads 
\begin{equation}
\frac{D\left(n_{\alpha}U^{\alpha}\right)}{d\lambda}=n_{\beta}a^{\beta}+2S_{\sigma\gamma\alpha}U^{\sigma}U^{\alpha}n^{\gamma}\,,\label{eq:Dervivative_naUa}
\end{equation}
where we have defined the acceleration vector appearing for non geodesics
fiducial curves as 
\begin{equation}
a^{\alpha}\equiv U^{\gamma}\nabla_{\gamma}U^{\alpha}\,.\label{eq:acceleration}
\end{equation}
The expression \eqref{eq:Dervivative_naUa} represents the failure
of the separation vector $n^{\alpha}$ and the tangent vector $U^{\alpha}$
to stay orthogonal to each other, that is, if at a given point $n^{\alpha}$
and $U^{\alpha}$ are orthogonal to each other, a general non-null
torsion, $S_{\alpha\beta\gamma}$, or a non-null acceleration, $a^{\alpha}$,
will spoil the preservation of such orthogonality along the curve.
Note that a torsion field, with no further imposed symmetry, will
lead to effects parallel to the direction of $U^{\alpha}$ (second
term on the right hand side of \eqref{eq:Dervivative_naUa}), contributing
to a relative acceleration between two initially infinitesimally close
particles.

The analysis of \eqref{eq:Dervivative_naUa} leads to the conclusion
that the tensor $B_{\alpha\beta}$ describing the behaviour of the
separation vector will have, for the case of a generic torsion tensor,
a non-null component tangential to the fiducial curve $c$. Without
loss of generality, it is then possible to write $B_{\alpha\beta}$
in terms of two components, one orthogonal and the other parallel
to $c$: given a projector $h_{\alpha\beta}$ onto the hypersurface
orthogonal to the curve $c$ at a given point, we can write 
\begin{equation}
B_{\alpha\beta}=B_{\perp\alpha\beta}+B_{\parallel\alpha\beta}\,,\label{eq:Bab_orth_tang_decomposition}
\end{equation}
with 
\begin{eqnarray}
B_{\perp\alpha\beta} & \equiv & h_{\alpha}^{\gamma}h_{\beta}^{\sigma}B_{\gamma\sigma}\,,\label{eq:Bab_orth_definition}\\
B_{\parallel\alpha\beta} & \equiv & B_{\alpha\beta}-B_{\perp\alpha\beta}\,.\label{eq:Bab_tang_definition}
\end{eqnarray}
In analogy to the General Relativity case, we want to define the kinematical
quantities identifying expansion, shear, and vorticity - $\theta$,
$\sigma_{\alpha\beta}$ and $\omega_{\alpha\beta}$, respectively
- of neighbouring curves of the congruence. These quantities will
only depend on the orthogonal part of the tensor $B_{\alpha\beta}$,
$B_{\perp\alpha\beta}$, so that, defining the expansion, shear and
vorticity as 
\begin{eqnarray}
\theta & = & B_{\perp\gamma}{}^{\gamma}\,,\label{eq:expansion_general}\\
\sigma_{\alpha\beta} & = & B_{\perp\left(\alpha\beta\right)}-\frac{h_{\alpha\beta}}{h_{\,\gamma}^{\gamma}}\theta\,,\\
\omega_{\alpha\beta} & = & B_{\perp\left[\alpha\beta\right]}\,,\label{eq:vorticity_general}
\end{eqnarray}
$B_{\perp\alpha\beta}$ can be decomposed into 
\begin{equation}
B_{\perp\alpha\beta}=\frac{h_{\alpha\beta}}{h_{\,\gamma}^{\gamma}}\theta+\sigma_{\alpha\beta}+\omega_{\alpha\beta}\,.\label{eq:B_orthl_esv_general}
\end{equation}

Before we continue and derive the Raychaudhuri equation let us emphasize
that the definitions of the kinematical quantities given by \eqref{eq:expansion_general}-\eqref{eq:vorticity_general}
are always valid whenever the tensor $B_{\alpha\beta}$ is related
with the variation of the separation vector between curves of a congruence
by an equation such as \eqref{eq:derivative_n_relation}. A formal
proof for this assertion can be seen in \cite{Poisson}.

\subsection{Raychaudhuri equation for a congruence of time-like curves}

The results in the previous section are quite general and valid for
curves of any kind; however, the procedure that defines the projector
$h_{\alpha\beta}$ strictly depends on the specific family of curves
considered. Once the projector is assigned, \eqref{eq:B_orthl_esv_general}
will give an actual expression in terms of the tangent vector and
the torsion tensor.

Let us then start by considering the generalized Raychaudhuri equation
for a congruence of time-like curves. In this case we will impose
that the fiducial curve is parametrised by the proper time, $\tau$,
and, in order to avoid confusion with the general case, we will label
its tangent vector as $v^{\alpha}$, with $v_{\alpha}v^{\alpha}=-1$.
The operator projecting onto the hypersurface orthogonal to $v^{\alpha}$
is given by 
\begin{equation}
h_{\alpha\beta}\equiv g_{\alpha\beta}+v_{\alpha}v_{\beta}\,,\label{eq:orthogonal_projector_timelike}
\end{equation}
and fulfills the following conditions 
\begin{equation}
\begin{alignedat}{1}h_{\alpha\beta}v^{\alpha} & =0\,,\\
h_{\alpha}^{\gamma}h_{\gamma\beta} & =h_{\alpha\beta}\,,\\
h_{\sigma}^{\sigma}= & N-1\,,
\end{alignedat}
\label{eq:timelike_projector_properties}
\end{equation}
where $N$ is, again, the dimension of the space-time.

In order to calculate the two components, orthogonal and parallel,
of $B_{\alpha\beta}$ another ingredient is necessary: we must find
an expression for the trajectories along which free particles move.
In a general non-Riemannian manifold, it is not an easy task to determine
such physical curves. The statement usually claimed in the literature
that particles follow geodesics, either of the Levi-Civita generic
connection or of the metric connection, turns out to be rather unsatisfactory
and \emph{naïve}. In a manifold equipped with a completely generic
affine connection, the particle trajectories can be correctly determined
starting from the equations of motion, with the latter obtained themselves
from the conservation laws for canonical energy-momentum and hypermomentum.
A comprehensive treatment for the calculation of the propagation equations
of single-pole and pole-dipole particles in metric-affine theories
of gravity has been developed in \cite{Puetzfeld}\footnote{We warn about a slight different notation in \cite{Puetzfeld}: for
example the definition of the torsion tensor is there twice our definition.} (see also \cite{Yasskin,Nomura} for early single-pole approximation
descriptions). A quite interesting result is that single-pole particles
without intrinsic hypermomentum follow, as in general relativity,
geodesics of the metric connection, $v^{\beta}\nabla_{\beta}^{(g)}v^{\alpha}=0$,
no matter what is the underlying theory of gravity.

Things become much more complicated for particles with non-vanishing
intrinsic hypermomentum. We will focus on the case of Riemann-Cartan
space-time, since in our setup nonmetricity is trivially zero. We
will also consider particles endowed with only mass and intrinsic
spin but we neglect dilaton charge and intrinsic shear for simplifying
reason. This means that the hypermomentum tensor reduces (for a single-pole
particle \cite{Puetzfeld}) to $\Delta_{[\mu\nu]}{}^{\rho}=\tau_{\mu\nu}v^{\rho}$,
where $\tau_{\mu\nu}$ is the anti-symmetric spin density tensor.
In this case the equations of motion for a particle read 
\begin{equation}
\begin{aligned}v^{\kappa}\nabla_{\kappa}v_{\alpha}= & -v_{\alpha}v^{\kappa}\nabla_{\kappa}\ln m-\frac{2}{m}\left[v^{\kappa}\nabla_{\kappa}\left(v^{\beta}v^{\sigma}\nabla_{\sigma}\tau_{\beta\alpha}\right)+\right.\\
 & +S_{\alpha\beta}{}^{\mu}v^{\beta}\left(mv_{\mu}+2v^{\nu}v^{\sigma}\nabla_{\sigma}\tau_{\nu\mu}\right)+\\
 & \left.+\frac{1}{2}R_{\alpha\beta\mu}{}^{\nu}v^{\beta}\tau^{\mu}{}_{\nu}\right]\equiv a_{\alpha}\,,
\end{aligned}
\end{equation}
\begin{equation}
v^{\alpha}\nabla_{\alpha}\tau_{\mu\nu}-2v_{[\mu}v^{\sigma}v^{\kappa}\nabla_{\kappa}\tau_{\sigma|\nu]}=0\,,\label{eq:ptau}
\end{equation}
Notably, this equation is independent by the specific choice of the
gravitational part in the action. It is important to stress that for
a non-spinning particle, $\tau_{\mu\nu}$ vanishes, the second of
equations \eqref{eq:ptau} is trivially satisfied while the first
one recovers the equation of the geodesics of the metric, $v^{\beta}\nabla_{\beta}^{(g)}v^{\alpha}=0$.
Note that the rest mass, that is the projection of the particle 4-momentum
on the rest frame, is guaranteed to be constant along the congruence
only in the case of non-spinning particle.

Using equations \eqref{eq:orthogonal_projector_timelike}-\eqref{eq:ptau}
in equations \eqref{eq:Bab_orth_definition} and \eqref{eq:Bab_tang_definition}
we find 
\begin{eqnarray}
B_{\perp\alpha\beta} & = & \nabla_{\alpha}v_{\beta}+2S_{\rho\alpha\beta}v^{\rho}+2S_{\rho\alpha\sigma}v^{\rho}v^{\sigma}v_{\beta}+v_{\alpha}a_{\beta}\,,\label{eq:Bab_ort_timelike}\\
B_{\parallel\alpha\beta} & = & -2S_{\rho\alpha\sigma}v^{\rho}v^{\sigma}v_{\beta}-v_{\alpha}a_{\beta}\,.
\end{eqnarray}
For the case of a congruence of time-like curves in a $N$-dimensional
space-time \eqref{eq:B_orthl_esv_general} can be written as 
\begin{equation}
B_{\perp\alpha\beta}=\frac{1}{N-1}h_{\alpha\beta}\theta+\sigma_{\alpha\beta}+\omega_{\alpha\beta}\,,\label{eq:B_orthl_esv_timelike}
\end{equation}
where \eqref{eq:timelike_projector_properties} was used.

Taking the variation of the tensor $B_{\perp\alpha\beta}$ along the
fiducial curve, and remembering the expression \eqref{eq:Riemann_tensor_definition}
for the Riemann tensor, we find 
\begin{equation}
\begin{aligned}\frac{D\,\,B_{\perp\alpha\beta}}{d\tau}= & v^{\gamma}\nabla_{\gamma}B_{\perp\alpha\beta}=R_{\gamma\alpha\beta\rho}v^{\rho}v^{\gamma}+\nabla_{\alpha}a_{\beta}-\\
 & -\left(\nabla_{\alpha}v^{\gamma}\right)\left(\nabla_{\gamma}v_{\beta}\right)-2v^{\gamma}S_{\gamma\alpha}\,^{\rho}\nabla_{\rho}v_{\beta}+\\
 & +2v^{\gamma}\nabla_{\gamma}\left(S_{\rho\alpha\beta}v^{\rho}\right)+v^{\gamma}\nabla_{\gamma}\left(v_{\alpha}a_{\beta}\right)+\\
 & +2v^{\gamma}\nabla_{\gamma}\left(S_{\rho\alpha\sigma}v^{\rho}v^{\sigma}v_{\beta}\right)\,,
\end{aligned}
\end{equation}
Shear and vorticity are, respectively, traceless and anti-symmetric,
so that, contracting $\alpha$ and $\beta$ on both sides of previous
equations, we obtain 
\begin{align}
\frac{D\theta}{d\tau}= & -R_{\gamma\rho}v^{\rho}v^{\gamma}+\nabla_{\alpha}a^{\alpha}-\left(\nabla_{\alpha}v^{\gamma}\right)\left(\nabla_{\gamma}v^{\alpha}\right)\nonumber \\
 & -2v^{\gamma}S_{\gamma\alpha}\,^{\rho}\nabla_{\rho}v^{\alpha}+2v^{\gamma}\nabla_{\gamma}\left(S_{\rho}v^{\rho}\right)\,,\label{eq:d_theta}
\end{align}
where $S_{\rho}\equiv S_{\rho\alpha}\,^{\alpha}$. Taking into account
that 
\begin{equation}
\begin{aligned}B_{\perp\alpha\beta}B_{\perp}^{\beta\alpha}= & \left(\nabla_{\alpha}v^{\beta}\right)\left(\nabla_{\beta}v^{\alpha}\right)+2v^{\rho}S_{\rho\beta}\,^{\alpha}\nabla_{\alpha}v^{\beta}+\\
 & +2S_{\rho\alpha}\,^{\beta}v^{\rho}\nabla_{\beta}v^{\alpha}+4S_{\rho\alpha}\,^{\beta}S_{\gamma\beta}\,^{\alpha}v^{\rho}v^{\gamma}+\\
 & +4S_{\alpha\beta\gamma}v^{\alpha}v^{\gamma}a^{\beta}\,,
\end{aligned}
\label{eq:BB}
\end{equation}
and using \eqref{eq:B_orthl_esv_timelike} we can write \eqref{eq:d_theta}
as 
\begin{equation}
\begin{aligned}\frac{D\theta}{d\tau}= & -R_{\gamma\rho}v^{\rho}v^{\gamma}-\left[\frac{1}{N-1}\theta^{2}+\sigma_{\alpha\beta}\sigma^{\alpha\beta}+\omega_{\alpha\beta}\omega^{\beta\alpha}\right]+\\
 & +2S_{\rho\alpha}\,^{\beta}v^{\rho}\left[\frac{1}{N-1}h_{\beta}\,^{\alpha}\theta+\sigma_{\beta}\,^{\alpha}+\omega_{\beta}\,^{\alpha}\right]\\
 & +\nabla_{\alpha}a^{\alpha}+2v^{\gamma}\nabla_{\gamma}\left(S_{\rho}v^{\rho}\right)+2S_{\rho\alpha}\,^{\beta}v^{\rho}v_{\beta}a^{\alpha}\,.
\end{aligned}
\label{eq:Raychaudhuri_timelike}
\end{equation}
This equation represents the generalization of the Raychaudhuri equation
for a time-like congruence of curves in the presence a generic torsion
field. Here, we would like to make a couple of comments: first of
all, this equation has been obtained using only geometrical arguments,
plus the canonical energy-momentum conservation equation to define
the equations of motion of the particles; this means that the result
is independent by the specific geometrical theory that we are choosing:
once that the theory has been assigned, then the Ricci tensor can
be related with the energy-momentum tensor accordingly with the (modified)
Einstein field equations. Secondly, it in interesting to stress that
the extra-force responsible of the acceleration term reported in the
first line of \eqref{eq:Raychaudhuri_timelike} is of purely geometric
origin, related to the extra coupling of the intrinsic spin (\emph{viz.}
intrinsic hypermomentum in the most general setup) with post-Riemannian
structures. Note also that our result differs from previous versions
of the torsion-full Raychaudhuri equation available in literature.
More concretely, refs.~\cite{Capozziello,Sengupta,Wanas} did not
take properly into account the relation between the tensor $B_{\alpha\beta}$
and the evolution of the separation vector between infinitesimally
close curves of the congruence, assuming that the expression of the
tensor $B_{\alpha\beta}$ is\emph{ a priori} the same as in the case
of null torsion. Few quite specific models for torsion accidentally
result anyway in the correct expression: retracing the properties
of the intrinsic spin, Refs.~\cite{Esposito,Tafel} consider the
simplifying assumption on the torsion tensor $S_{\alpha\beta}\,^{\gamma}=S_{\alpha\beta}v^{\gamma}$,
with $S_{\left(\alpha\beta\right)}=0$ and $S_{\alpha\beta}v^{\alpha}=0$;
in Ref.~\cite{Stewart} instead torsion is constrained to be $S_{\alpha\beta}\,^{\gamma}=\eta_{\alpha\beta\sigma\epsilon}\,v^{\gamma}v^{\sigma}S^{\epsilon}$,
where $\eta_{\alpha\beta\sigma\epsilon}$ is the completely anti-symmetric
Levi-Civita tensor. However, in these special cases the extra symmetries
imposed on the torsion tensor imply that the second term on the right
hand side of \eqref{eq:B_tensor_general} is null, reducing the problem
to the torsion free case.

\subsection{Raychaudhuri equation for a congruence of null curves\label{subsec:Raychaudhuri_null_congruencel}}

Let us now derive the Raychaudhuri equation for a congruence of null
curves. As in the previous subsection, to avoid any confusion, we
will re-label the tangent vector to the fiducial curve and call it
$k^{\alpha}$, so that $k^{\alpha}k_{\alpha}=0$.

In this case, unfortunately, if $\widetilde{h}_{\alpha\beta}$ is
the projector onto the hypersurface that is orthogonal to the fiducial
null curve, it cannot be naively defined by \eqref{eq:orthogonal_projector_timelike},
since it would not be orthogonal to $k^{\alpha}$ (it would be $\widetilde{h}_{\alpha\beta}k^{\alpha}=k_{\beta}\neq0$).
The way out of this problem is through the introduction of an auxiliary
null vector field, $\xi^{\alpha}$, such that \cite{Carter} 
\begin{eqnarray}
k^{\alpha}\xi_{\alpha} & = & -1\,,\label{eq:xi_ka_dot_product}\\
\xi_{\alpha}\xi^{\alpha} & = & 0\,.\label{eq:xi_na_orth_condition}
\end{eqnarray}
Using \eqref{eq:xi_ka_dot_product} and \eqref{eq:xi_na_orth_condition}
we can now properly introduce a projector onto the hypersurface orthogonal
to both $k^{\alpha}$ and $\xi^{\alpha}$ as 
\begin{equation}
\widetilde{h}_{\alpha\beta}=g_{\alpha\beta}+k_{\alpha}\xi_{\beta}+\xi_{\alpha}k_{\beta}\,,\label{eq:null_projector}
\end{equation}
satisfying the following properties 
\begin{equation}
\begin{alignedat}{1}\widetilde{h}_{\alpha\beta}k^{\alpha}= & \widetilde{h}_{\alpha\beta}\xi^{\alpha}=0\,,\\
\widetilde{h}_{\,\alpha}^{\sigma}\widetilde{h}_{\sigma\beta}= & \widetilde{h}_{\alpha\beta}\,,\\
\widetilde{h}_{\sigma}^{\sigma}= & N-2\,.
\end{alignedat}
\label{eq:null_projector_properties}
\end{equation}

As in the time-like case, the extra ingredient to be taken into account
is the effective particles trajectories. An important {\em caveat}
is here due: particles that are moving along null curves are massless
particle; the most natural candidates in the Standard Model are then
photons. However, the minimal coupling procedure to generalize the
electromagnetic field to non-Riemannian environments preserve photons
by having a non-vanishing intrinsic hypermomentum; hence, photons
follow the geodesics determined by the Christoffel symbols, 
\begin{equation}
k^{\alpha}\nabla_{\alpha}^{(g)}k^{\beta}=0\,.\label{geo2}
\end{equation}
As a side note, let us stress that some theories, such as the Standard
Model Extension \cite{Colladay}, allow for a non-minimal coupling
of the electromagnetic field with geometry (and eventually other fields),
which means a non trivial intrinsic hypermomentum: anyway, in this
case the extra operators will also introduce some effective mass for
the photon (the simplest case being the explicitly massive Proca field)
and \emph{a fortiori} they could not follow null curves.

In the context of the Standard Model, neutrinos deserve a separate
mention; the recent confirmation of the phenomenon of neutrino oscillations
\cite{Fukuda,Ahmad} directly imply that neutrinos might be massive
\cite{Bilenky}, avoiding them to follow null paths\footnote{For the sake of completeness, we want to mention that also in this
case there is still room for the (unlikely) scenario with one out
of three neutrino species exactly massless.}. In spite of that, let us take into account the behaviour of the
Standard Model massless neutrino, \emph{viz.} a single-pole, massless
Dirac particle. Dirac particles have a completely anti-symmetric hypermomentum,
which implies, in the single-pole approximation, a vanishing intrinsic
spin density tensor, $\tau_{\mu\nu}=0$ \cite{Blago,Vojinovic}. Back
to \eqref{eq:ptau}, this would mean that the particles follow metric
geodesics, as in the non-spinning case, and as for photons.

It is finally possible to use \eqref{eq:xi_ka_dot_product}-\eqref{geo2}
to find the orthogonal and tangential components of the tensor $B_{\alpha\beta}$
that defines the dynamics of the separation vector of the null congruence
of geodesics; a straightforward calculation gives

\begin{align}
B_{\perp\alpha\beta} & =\nabla_{\alpha}k_{\beta}-2S_{\alpha\gamma\beta}k^{\gamma}-2S_{\alpha\gamma\sigma}k^{\gamma}k^{\sigma}\xi_{\beta}+2S_{\gamma\sigma\beta}k_{\alpha}k^{\gamma}\xi^{\sigma}\nonumber \\
 & +2S_{\gamma\rho\sigma}k_{\alpha}k^{\gamma}k^{\sigma}\xi_{\beta}\xi^{\rho}+k_{\alpha}\xi^{\gamma}\nabla_{\gamma}k_{\beta}+k_{\beta}\xi^{\gamma}\nabla_{\alpha}k_{\gamma}\nonumber \\
 & -2S_{\alpha\gamma\sigma}k_{\beta}k^{\gamma}\xi^{\sigma}+k_{\alpha}k_{\beta}\xi^{\gamma}\xi^{\sigma}\nabla_{\sigma}k_{\gamma}\nonumber \\
 & +2S_{\beta\gamma\sigma}k^{\gamma}k^{\sigma}\xi_{\alpha}+2S_{\gamma\sigma\rho}k_{\alpha}k_{\beta}k^{\gamma}\xi^{\rho}\xi^{\sigma}\nonumber \\
 & +2k_{\beta}\xi_{\alpha}\xi^{\sigma}S_{\sigma\gamma\rho}k^{\gamma}k^{\rho}\label{eq:Bab_ort_null}
\end{align}
\begin{align}
B_{\parallel\alpha\beta} & =-2S_{\rho\sigma\beta}k^{\rho}k_{\alpha}\xi^{\sigma}-k_{\beta}\xi^{\rho}\nabla_{\alpha}k_{\rho}+2S_{\alpha\rho\sigma}k^{\rho}k_{\beta}\xi^{\sigma}\nonumber \\
 & -2S_{\rho\sigma\gamma}k^{\rho}k_{\alpha}k_{\beta}\xi^{\sigma}\xi^{\gamma}+2S_{\alpha\rho\sigma}k^{\rho}k^{\sigma}\xi_{\beta}\nonumber \\
 & -2S_{\rho\gamma\sigma}k^{\rho}k^{\sigma}k_{\alpha}\xi^{\gamma}\xi_{\beta}-2\xi^{\sigma}\xi_{\alpha}k_{\beta}S_{\sigma\gamma\rho}k^{\gamma}k^{\rho}\nonumber \\
 & -2\xi_{\alpha}S_{\beta\gamma\sigma}k^{\gamma}k^{\sigma}-k_{\alpha}\xi^{\rho}\nabla_{\rho}k_{\beta}\nonumber \\
 & -k_{\alpha}k_{\beta}\xi^{\rho}\xi^{\sigma}\nabla_{\sigma}k_{\rho}
\end{align}

Similarly to what was done in \eqref{eq:B_orthl_esv_general}, we
can relate the component $B_{\perp\alpha\beta}$ with the expansion,
shear and vorticity of the congruence of the null curves, 
\begin{equation}
B_{\perp\alpha\beta}=\frac{\widetilde{h}_{\alpha\beta}}{N-2}\theta+\sigma_{\alpha\beta}+\omega_{\alpha\beta}\,,\label{eq:B_orthl_esv_null}
\end{equation}
that in combination with \eqref{eq:Bab_ort_null} provides an expression
for the expansion of the congruence of null geodesics in terms of
$k^{\alpha}$, $\xi^{\alpha}$ and $S_{\alpha\beta\gamma}$, given
by 
\begin{eqnarray}
\theta & = & \nabla_{\gamma}k^{\gamma}+2S_{\gamma}k^{\gamma}=\nabla_{\gamma}^{(g)}k^{\gamma}\equiv\widetilde{\theta}\,.\label{eq:expansion_equivalence}
\end{eqnarray}
Note
that also in this case, as in general relativity, the scalar expansion
$\theta$ does not depend on the auxiliary vector $\xi^{\mu}$ chosen
to define a proper projection operator to treat the null curves congruence case.

\textcolor{black}{Tracing Eq.~\eqref{eq:Bab_ort_null}
and computing its covariant derivative along the fiducial null curve
we find}
\begin{equation}
\begin{aligned}\frac{D\theta}{d\lambda}& =-R_{\alpha\beta}k^{\alpha}k^{\beta}-\left(\frac{1}{N-2}\theta^{2}+\sigma_{\alpha\beta}\sigma^{\alpha\beta}+\omega_{\alpha\beta}\omega^{\beta\alpha}\right)\\
 & +2S_{\rho\alpha}{}^{\beta}k^{\rho}\left(\frac{\tilde{h}_{\beta}{}^{\alpha}}{N-2}\theta+\sigma_{\beta}{}^{\alpha}+\omega_{\beta}{}^{\alpha}\right)+2k^{\gamma}\nabla_{\gamma}\left(S_{\rho}k^{\rho}\right)\\
 & +2\nabla_{\alpha}\left(S^{\alpha}\,_{\gamma\rho}k^{\gamma}k^{\rho}\right)+4\left(h^{\beta\gamma}S_{\alpha\beta\gamma}k^{\alpha}-S_{\alpha}k^{\alpha}\right)^{2}\\
 & +4S_{\mu}{}^{\alpha\beta}k^{\mu}\left[S_{\delta\beta\alpha}k^{\delta}-h_{\beta}^{\rho}S_{\delta\rho\alpha}k^{\delta}\right]\\
 & +S^{\alpha}\,_{\mu\nu}k^{\mu}\left(\widetilde{h}^{\nu\beta}-g^{\nu\beta}\right)\left(4B_{\parallel\alpha\beta}+6B_{\parallel\beta\alpha}\right)\\
 & -2B_{\parallel\delta}{}^{\alpha}B_{\parallel\beta\alpha}\left(\widetilde{h}^{\delta\beta}-g^{\delta\beta}\right)\,,
\end{aligned}
\label{eq:Ray_null}
\end{equation}
which represents the Raychaudhuri equation for a
congruence of null geodesics in the presence of a torsion tensor field.
Here the literature is more sparse: ref.~\cite{Visser} derives the
Raychaudhuri equation for a null congruence of curves in the particular
case of a completely anti-symmetric torsion, missing in any case the
correct definition of the $B_{\alpha\beta}$ tensor (although without affecting the final result).

Looking at Eq.~\eqref{eq:Ray_null} it is not
clear that the Raychaudhuri equation for a null congruence is independent
of the choice of the auxiliary null vector $\xi^{\alpha}$ that was
introduced to define the projector onto the orthogonal hypersurface
to the congruence, Eq.~\eqref{eq:null_projector}: we have
terms that depend on the projector $\widetilde{h}^{\alpha\beta}$
and $B_{\parallel\alpha\beta}$, which themselves depend on $\xi^{\alpha}$.
However, since the expansion $\theta$ is a scalar quantity, its rate
of variation along the congruence will be related to the the rate
of variation of $\widetilde{\theta}$ by 
\begin{equation}
\frac{D\theta}{d\lambda}=k^{\mu}\nabla_{\mu}\theta=k^{\mu}\partial_{\mu}\theta=k^{\mu}\partial_{\mu}\widetilde{\theta}=k^{\mu}\nabla_{\mu}^{(g)}\widetilde{\theta}=\frac{\widetilde{D}\widetilde{\theta}}{d\lambda}\,.
\end{equation}
where $\frac{\widetilde{D}}{d\lambda}=k^{\mu}\nabla_{\mu}^{(g)}$
represents the covariant derivative along the fiducial curve where
only the metric connection is considered. Computing
the derivative, we recover the usual general relativity expression
for the Raychaudhuri equation governing the evolution of the expansion:

\begin{equation}
\frac{\widetilde{D}\widetilde{\theta}}{d\lambda}=-\widetilde{R}_{\alpha\beta}k^{\alpha}k^{\beta}-\left[\frac{1}{N-2}\widetilde{\theta}^{2}+\widetilde{\sigma}_{\alpha\beta}\widetilde{\sigma}^{\alpha\beta}+\widetilde{\omega}_{\alpha\beta}\widetilde{\omega}^{\beta\alpha}\right]\,,\label{eq:Ray_null_final}
\end{equation}
where the tilded quantities are calculated with the Christoffel connection,
and 
\begin{equation}
\begin{aligned}\widetilde{\theta}= & \nabla_{\mu}^{(g)}k^{\mu}\,,\\
\widetilde{\sigma}_{\mu\nu}= & \nabla_{(\mu}^{(g)}k_{\nu)}+\xi^{\gamma}k_{(\mu}\nabla_{\gamma}^{(g)}k_{\nu)}+\xi^{\gamma}k_{(\nu}\nabla_{\mu)}^{(g)}k_{\gamma}\\
 & +k_{\mu}k_{\nu}\xi^{\gamma}\xi^{\sigma}\nabla_{\sigma}^{(g)}k_{\gamma}-\nabla_{\mu}^{(g)}k^{\mu}\,,\\
\widetilde{\omega}_{\mu\nu}= & \nabla_{[\mu}^{(g)}k_{\nu]}+k_{[\mu}\xi^{\gamma}\nabla_{\gamma}^{(g)}k_{\nu]}+\xi^{\gamma}k_{[\nu}\nabla_{\mu]}^{(g)}k_{\gamma}\,.
\end{aligned}
\end{equation}
Note that, in spite of the explicit appearance of the auxiliary vector
$\xi_{\mu}$ in the two expressions for $\widetilde{\sigma}_{\mu\nu}$
and $\widetilde{\omega}_{\mu\nu}$, the total expression \eqref{eq:Ray_null_final}
does not depend on $\xi_{\mu}$, as becomes evident after calculating the contractions
\begin{equation}
\begin{aligned}\widetilde{\sigma}_{\mu\nu}\widetilde{\sigma}^{\mu\nu}= & -\frac{1}{N-2}\tilde{\theta}^{2}+\frac{2}{N-2}\tilde{\theta}+\frac{1}{2}\nabla_{\mu}^{(g)}k^{\nu}\cdot\nabla_{\nu}^{(g)}k^{\mu}\\
 & +\frac{1}{2}\nabla_{\mu}^{(g)}k^{\nu}\cdot\nabla^{(g)}{}^{\mu}k_{\nu}\,,\\
\widetilde{\omega}_{\mu\nu}\widetilde{\omega}^{\nu\mu}= & \frac{1}{2}\nabla_{\mu}^{(g)}k^{\nu}\cdot\nabla_{\nu}^{(g)}k^{\mu}-\frac{1}{2}\nabla_{\mu}^{(g)}k^{\nu}\cdot\nabla^{(g)}{}^{\mu}k_{\nu}\,.
\end{aligned}
\end{equation}
Since Eq.~\eqref{eq:Ray_null_final} is equivalent to Eq.~\eqref{eq:Ray_null},
we find that the Raychaudhuri equation for a null congruence is independent
of the vector field $\xi^{\mu}$.

As a final comment, let us address the following important remark:
one could be tempted to naively think that, since \eqref{eq:Ray_null_final}
does not depend explicitly of the torsion tensor, then the evolution
of the expansion of the curves followed by massless particles does
not depend on the presence of a torsion field. This conclusion is
not correct since although the torsion tensor does not explicitly
appear in \eqref{eq:Ray_null_final}, it does affect the geometry
of the space-time in a way described by the (modified) Einstein field
equations; the metric solution of the Einstein equations will be itself
affected by the such modification\footnote{Just as a workable example, consider the simple Einstein–Cartan case:
one of the equations of motion describes algebraically the torsion
tensor as a function of the spin tensor; the other equation, rewritten
in terms of the Christoffel connection eliminating the torsion tensor,
relates the (Christoffel) Einstein tensor to a combined version of
the stress-energy tensor, that now takes into account also spin density
terms \cite{Hehl}.}, and so will be the corresponding (metric) Ricci tensor $\widetilde{R}_{\mu\nu}$
appearing in \eqref{eq:Ray_null_final}.

\section{Conclusions \label{sec:Conclusion}}

In this paper we derived the equation for the evolution of the separation
vector between infinitesimally close curves of a congruence in space-times
with non-null generic torsion field, clarifying some of the ambiguities
lingering in the literature about the role of the torsion tensor.
We concluded that the presence of a torsion field leads in general
to tangent and orthogonal effects to the congruence, in particular,
the presence of a generic torsion field contributes to a relative
acceleration between test particles. This effects happen either for
free-falling or accelerated particles following time-like, null or
space-like curves.

The evolution equation of the separation vector can be further separated
and be used to study the kinematical quantities that characterize
a congruence of curves, namely the expansion, shear and vorticity.
We derived, for the first time in the literature, how such kinematical
quantities depend on a completely generic torsion field.

Knowing how the kinematical decomposition of the geodesics congruence
is influenced by the torsion tensor allows the possibility to test
models equipped with a nontrivial Riemannian connection through the
study of the motion of test particles. Let us expand a bit on this
point: the matter source appearing on the right hand side of the equation
of motion for the torsion tensor (what is usually dubbed hypermomentum,
obtained from the variation of the classical matter action with respect
to the independent connection) depends on the specific coupling between
matter and torsion itself: different models will lead to different
field equations for the torsion and henceforth to different solutions.
This means that the eventual contributions of torsion to the various
kinematical quantities and their evolution will be dependent on the
specific chosen model; the analysis then of the evolution of a matter
fluid in a region of space-time will give a chance to distinguish
between different allowed couplings.

One of the most important achievements of this paper is the generalization
of Raychaudhuri equation - the equation for the evolution of the expansion
of a congruence of curves - for the case of time-like and null curves
in an $N$-dimensional space-time with the most generic torsion field.
The study of the evolution of the expansion of time-like and null
curves in spacetime is obviously important due to its role in defining
the evolution of gravitational collapse and possible formation of
singularities. Moreover, as was shown in Ref.~\cite{Visser}, the
expansion of null light-rays also plays a preponderant role on the
definition of the throats of (dynamical) wormholes, namely exotic
solutions requiring the violation of energy conditions; a still open
question is whether or not the degrees of freedom of a completely
generic torsion field can somehow avoid the violation of the \emph{null
energy condition} of whatever matter present at the dynamical wormhole's
throat. We firmly believe that the study of such possibilities will
be of great interest in the near future.\\

\begin{acknowledgments}
The Authors wish to thank José P. S. Lemos for early discussions on
a first version of the paper. We thank FCT-Portugal for financial
support through Project No. PEst - OE/FIS/UI0099/ 2015. PL thanks
IDPASC and FCT-Portugal for financial support through Grant No. PD/BD/114074/2015.
VV is supported by the FCT-Portugal grant SFRH/BPD/77678/2011. 
\end{acknowledgments}


\begin{thebibliography}{10}
\bibitem{Raychaudhuri}A. K. Raychaudhuri, ``Relativistic Cosmology
I'', Phys. Rev. \textbf{98}, 1123 (1955).

\bibitem{Penrose}R. Penrose, ``Gravitational Collapse and Space-Time
Singularities'', Phys. Rev. Lett. \textbf{14}, 57 (1965).

\bibitem{Hawking}S. Hawking and G. F. R. Ellis, \emph{The Large Scale
Structure of Space-Time} (Cambridge University Press, Cambridge, 1973).

\bibitem{Sengupta} S. Kar and S. Sengupta, ``The Raychaudhuri equations:
A brief review'', Pramana \textbf{69}, 49 (2007).

\bibitem{Tomita}K. Tomita, ``Gravitational lens effect of wall-like
objects and its cosmological implications'', Prog. Theor. Phys. \textbf{85},
57 (1991).

\bibitem{herrera}L. Herrera, ``Cracking of self-gravitating compact
objects'', Phys. Lett. A \textbf{165}, 206 (1992).

\bibitem{Jacobson}T. Jacobson, ``Thermodynamics of spacetime: The
Einstein equation of state'', Phys. Rev. Lett. \textbf{75}, 1260
(1995).

\bibitem{Eling_Guedens_Jacobson}C. Eling, R. Guedens and T. Jacobson,
``Nonequilibrium thermodynamics of spacetime'', Phys. Rev. Lett.
\textbf{96}, 121301 (2006).

\bibitem{chirco}G. Chirco and S. Liberati, ``Non-equilibrium Thermodynamics
of Spacetime: The Role of Gravitational Dissipation'', Phys. Rev.
D \textbf{81}, 024016 (2010).

\bibitem{Hehl}F. W. Hehl, P. von der Heyde, G. D. Kerlick and J.
M. Nester, ``General relativity with spin and torsion: Foundations
and prospects'', Rev. Mod. Phys. \textbf{48}, 393 (1976).

\bibitem{Trautman}A. Trautman, ``Spin and torsion may avert gravitational
singularities.'', Nature Phys. Sci. \textbf{242}, 7 (1973).

\bibitem{Stewart}J. Stewart and P. Hájíček, ``Can spin avert singularities?'',
Nature Phys. Sci. \textbf{244}, 96 (1973).

\bibitem{Kop}W. Kopczýnski, ``An anisotropic universe with torsion.'',
Phys. Lett. A \textbf{43A}, 63 (1973).

\bibitem{Tafel}J. Tafel, ``A non-singular homogeneous universe with
torsion'', Phys. Lett. A \textbf{45}, 4 (1973).

\bibitem{sotiriou} T. P. Sotiriou and S. Liberati, ``Metric-affine
f(R) theories of gravity,'' Annals Phys.\textbf{ 322}, 935 (2007).

\bibitem{enzo2} V. Vitagliano, T. P. Sotiriou and S. Liberati, ``The
dynamics of generalized Palatini Theories of Gravity,'' Phys.\ Rev.\ D
\textbf{82}, 084007 (2010).

\bibitem{enzo3} V. Vitagliano, T. P. Sotiriou and S. Liberati, ``The
dynamics of metric-affine gravity,'' Annals Phys.\textbf{ 326}, 1259
(2011) Erratum: {[}Annals Phys.\textbf{ 329}, 186 (2013){]}.
 
\bibitem{Santana}
  L.~T.~Santana, M.~O.~Calvão, R.~R.~R.~Reis and B.~B.~Siffert,
  ``How does light move in a generic metric-affine background?,''
  Phys.\ Rev.\ D \textbf{95},  061501 (2017).
  
\bibitem{Capozziello}Y. Cai, S. Capozziello, M. De Laurentis and
E. N. Saridakis, ``$f\left(T\right)$ teleparallel gravity and cosmology'',
Rept. Prog. Phys. \textbf{79}, 106901 (2016).

\bibitem{Mao}Y. Mao, M. Tegmark, A. H. Guth, and S. Cabi, ``Constraining
torsion with Gravity Probe B'', Phys. Rev. D \textbf{76}, 104029
(2007).

\bibitem{Enzo}V. Vitagliano, ``The role of nonmetricity in metric-affine
theories of gravity'', Class. Quant. Grav. \textbf{31}, 045006 (2014).

\bibitem{enzobis}V. Vitagliano, ``On the dynamical content of MAGs,''
J.\ Phys.\ Conf.\ Ser.\ \textbf{600}, no. 1, 012043 (2015).

\bibitem{Poisson}E. Poisson, \emph{A Relativist's Toolkit: The Mathematics
of Black-Hole Mechanics} (Cambridge University Press, Cambridge, 2004).

\bibitem{Puetzfeld} D.~Puetzfeld and Y.~N.~Obukhov, ``Propagation
equations for deformable test bodies with microstructure in extended
theories of gravity,'' Phys.\ Rev.\ D \textbf{76}, 084025 (2007)
Erratum: {[}Phys.\ Rev.\ D \textbf{79}, 069902 (2009){]}.

\bibitem{Yasskin} P.~B.~Yasskin and W.~R.~Stoeger, S.J., ``Propagation
Equations for Test Bodies With Spin and Rotation in Theories of Gravity
With Torsion,'' Phys.\ Rev.\ D \textbf{21}, 2081 (1980).

\bibitem{Nomura} K.~Nomura, T.~Shirafuji and K.~Hayashi, ``Semiclassical
particles with arbitrary spin in the Riemann-Cartan space-time,''
Prog.\ Theor.\ Phys.\ \textbf{87}, 1275 (1992).

\bibitem{Wanas}M. I. Wanas and M. A. Bakry, ``Effect of spin-torsion
interaction on Raychaudhuri equation'', Int. J. Mod. Phys. A \textbf{24},
5025 (2009).

\bibitem{Esposito}G. Esposito, ``Mathematical structures of space-time'',
Fortsch. Phys. \textbf{40}, 1 (1992).

\bibitem{Carter}B. Carter, ``The general theory of the mechanical,
electromagnetic and thermodynamic properties of black holes”, in General
Relativity: an Einstein Centenary Survey, ed. S.W. Hawking and W.
Israel (Cambridge University Press, Cambridge, England, 1979).

\bibitem{Colladay} D.~Colladay and V.~A.~Kostelecky, ``Lorentz
violating extension of the standard model,'' Phys.\ Rev.\ D \textbf{58},
116002 (1998).

\bibitem{Fukuda} Y.~Fukuda \textit{et al.} {[}Super-Kamiokande Collaboration{]},
``Evidence for oscillation of atmospheric neutrinos'', Phys.\ Rev.\ Lett.\ \textbf{81},
1562 (1998).

\bibitem{Ahmad} Q.~R.~Ahmad \textit{et al.} {[}SNO Collaboration{]},
``Direct evidence for neutrino flavor transformation from neutral
current interactions in the Sudbury Neutrino Observatory'', Phys.\ Rev.\ Lett.\ \textbf{89},
011301 (2002).

\bibitem{Bilenky} S.~M.~Bilenky and S.~T.~Petcov, ``Massive
Neutrinos and Neutrino Oscillations,'' Rev.\ Mod.\ Phys.\ \textbf{59},
671 (1987) Erratum: {[}Rev.\ Mod.\ Phys.\ \textbf{61}, 169 (1989){]}
Erratum: {[}Rev.\ Mod.\ Phys.\ \textbf{60}, 575 (1988){]}.

\bibitem{Blago} M.~Blagojević and F.~W.~Hehl, chapter 14 in \emph{Gauge
Theories of Gravitation : A Reader with Commentaries}, (Imperial College
Press, London, 2013).

\bibitem{Vojinovic} M.~Vasilic and M.~Vojinovic, ``Zero-size objects
in Riemann-Cartan spacetime,'' JHEP \textbf{0808}, 104 (2008).

\bibitem{Visser}D. Hochberg and M. Visser, ``Dynamic wormholes,
anti-trapped surfaces, and energy conditions'', Phys. Rev. D \textbf{58},
044021 (1998). 
\end{thebibliography}
\end{document}